\RequirePackage{fix-cm}
\documentclass{article}
\usepackage{graphicx}
\usepackage{epstopdf}
\usepackage[T1]{fontenc}
\usepackage[utf8]{inputenc}
\usepackage[english]{babel}
\newcommand{\maniv}{\mathcal{M}}
\newcommand{\mmaniv}{(\maniv,\g{}{ab})}
\newcommand{\maniii}{\Sigma}

\newcommand{\manii}{\mathcal{S}}

\newcommand{\ball}{\mathcal{B}}
\newcommand{\world}{\Omega}

\newcommand{\ro}{\varrho}
\newcommand{\teta}{\vartheta}
\newcommand{\fii}{\varphi}

\newcommand{\expn}{\theta^{(n)}}
\newcommand{\expl}{\theta^{(l)}}

\newcommand{\dd}{\mathrm{d}}
\newcommand{\DD}{\mathrm{D}}
\newcommand{\diff}{\partial}

\newcommand{\vol}[1]{\varepsilon_{#1}}

\newcommand{\g}[2]{{g^{#1}}_{#2}}
\newcommand{\h}[2]{{h^{#1}}_{#2}}

\newcommand{\TT}[2]{{T^{#1}}_{#2}}
\newcommand{\GG}[2]{{G^{#1}}_{#2}}
\newcommand{\FF}[2]{{F^{#1}}_{#2}}
\newcommand{\RR}[2]{{R^{#1}}_{#2}}
\begin{document}

\title{Geometric inequalities in spherically symmetric spacetimes}
\author{K\'{a}roly Csuk\'{a}s\thanks{e-mail: csukas.karoly@wigner.mta.hu}\\Wigner RCP,\\H-1121 Budapest, Konkoly Thege Mikl\'{o}s \'{u}t 29-33.\\Hungary}

\date{Received: date / Accepted: date}

\maketitle

\begin{abstract}
In geometric inequalities ADM mass plays more fundamental role than the concept of quasi-local mass. This paper is to demonstrate that using the quasi-local mass some new insights can be acquired. In spherically symmetric spacetimes the Misner-Sharp mass and the concept of the Kodama vector field provides an ideal setting to the investigations of geometric inequalities. We applying the proposed new techniques to investigate the spacetimes containing black hole or cosmological horizons but we shall also apply them in context of normal bodies. Most of the previous investigations applied only the quasi-local charges and the area. Our main point is to include the quasi-local mass in the corresponding geometrical inequalities. This way we recover some known relations but new inequalities are also derived.
\end{abstract}

\section{Introduction}
\label{sec:intro}
The study of non-stationary black holes is highly complicated. Even stationary black hole spacetime may contain complex structures \cite{Ansorg:2005bq}. However it seems that inequalities between stationary black hole's characteristic quantities may remain valid for dynamical black hole spacetime. The characteristic quantities are the spacetime's mass and angular momentum and the black hole's surface area and charge. These inequalities provide important insight into black hole evolution and may serve as basis to derive laws of black hole dynamics. Recently similar inequalities was obtained for bodies which provided criteria for black hole formation.\\
The authors in \cite{Hennig:2008yw,Hennig:2008zy} solve the Einstein-Maxwell equations in a vicinity of the black hole horizon using metric and electromagnetic potentials. They reformulate the problem as a variational problem. In this way they provide the (\ref{eq:AJQ}) inequality for arbitrary rotating, charged, axisymmetric and stationary black hole surrounded by matter.
\begin{equation}
\label{eq:AJQ}
 \left(\frac{8\pi J^2}{A}\right)^2+\left(\frac{4\pi Q^2}{A}\right)\leq 1
\end{equation}
This inequality was extended to non-stationary black holes without charge in \cite{Dain:2011pi,Jaramillo:2011pg}, then in \cite{Clement:2012vb} including the electromagnetic charge.\\
Inequalities including mass was provided in asymptotically flat spacetimes in \cite{Dain:2005qj}, then including charges in \cite{Chrusciel:2009ki}. The inequality involving the ADM-mass reads as:
\begin{equation}
 \label{eq:MQJ}
 M_{ADM}\geq \sqrt{\frac{J^2}{M^2_{ADM}}+Q^2}.
\end{equation}
Inequality (\ref{eq:MQJ}) was proven in axisymmetric, asymptotically flat, electrovacuum spacetimes. The authors in \cite{Ansorg:2007fh} proved that Christodoulou-Ruffini mass satisfies the same inequality. The Christodoulou-Ruffini mass is defined through the black hole's electric charge and angular momentum. In this way they get a fully quasi-local result for degenerate ($\kappa=0$) black holes.\\
The inequalities mentioned so far are valid for black holes, however in \cite{Dain:2011kb} the authors provided the inequality
\begin{equation}
 \label{eq:AQb}
 A\geq \frac{4\pi}{3} Q^2
\end{equation}
for surfaces in spacetimes without black holes. Furthermore in \cite{Khuri:2015zla,Khuri:2015xpa} it was shown that if (\ref{eq:AJQ}) is violated for a surface, the concentration of charge or angular momentum create a black hole.\\
In \cite{Reiris:2014tva} the author using the quasi-local notion of Hawking mass in spherically symmetric spacetime proves the inequality
\begin{equation}
\label{eq:reiris}
 \frac{\pi Q^4}{M^2}\leq A
\end{equation}
in locally Reissner-Nordström spacetime for electrically charged body. Further developing this argument in \cite{Anglada:2015tan} the authors provide a wider investigation for spherically symmetric asymptotically flat spacetimes. For charged body the
\begin{equation}
 2R>|Q|,
\end{equation}
and for black holes the inequality 
\begin{equation}
 R_{h}\geq |Q|
\end{equation}
was proved.\\
The Hawking mass refers to the null expansions which determines whether a surface is trapped or untrapped. This way inequalities may be deduced from the fact that a surface is marginally trapped or untrapped. In spherically symmetric spacetimes two important simplification occur: the area-radius provides a natural mean to measure the size of bodies and the Hawking mass reduces to Misner-Sharp mass which is also straightforwardly related to stress-energy tensor. By choosing the stress-energy tensor suitably we can easily construct physically interesting examples. By doing so we aim to extend the validity of former investigations by dropping the assumption of asymptotic flatness or any reference to Reissner-Nordström spacetime. We provide a generalization of these former result which may be applied to arbitrary spherically symmetric spacetimes in particular to the Bertotti-Robinson spacetime \cite{Bertotti:1959pf,Robinson:1959ev}.\\
We use this method to examine the following three systems. First a charged black hole modeled by a marginally outer trapped surface surrounded by electrovacuum in it's immediate neighborhood. Apart from this neighborhood any configuration of matter allowed. Second a charged body modeled by an untrapped surface surrounded by electrovacuum as before. This requirement concerning the electrovacuum region is to separate the body or black hole from the rest of the spacetime. The third example applies flat cosmology to demonstrate the usage for inner marginally trapped surfaces.\\
The paper is organized as follows. In section \ref{sec:prelim} we briefly introduce the notions which are used throughout the paper. In section \ref{sec:em} some facts are provided about electromagnetic matter in spherical symmetry which we will use in the following section. In section \ref{sec:app} several applications will be considered. In section \ref{sec:mass} the notions used in \ref{sec:msem} and \ref{sec:chb} are introduced. In section \ref{sec:msem} we derive the standard inequality between area and charge for black holes. It is worth mentioning that our method provides more: besides the usual inequalities between area and mass a set of inequalities applicable in case of future inner marginal surface. In \ref{sec:chb} the notion of normal charged body is introduced. Finally section \ref{sec:cosm} is to demonstrate that the future inner marginal surface comes with some important properties. Our results are summarized in section \ref{sec:summ}.

\section{Preliminaries}
\label{sec:prelim}

In this section the theory in which our results hold and the conventions used are introduced.

\subsection{Spacetime}
Spacetime is represented by a $\mmaniv$ pair, where $\maniv$ is a $4$-dimensional, smooth, paracompact, connected, orientable manifold with a smooth Lorentzian metric $\g{}{ab}$ with signature $(-,+,+,+)$. It is assumed that $\mmaniv$ is time orientable, spherically symmetric and Einstein's equation 
\begin{equation}
\GG{}{ab}=\RR{}{ab}-\frac{\g{}{ab}}{2}\RR{}{}=8\pi\TT{}{ab}
\end{equation}
holds. Everywhere in this paper $G_N=c=1$.

$\mmaniv$ is said to satisfy the dominant energy condition (DEC) if for all future directed timelike vector $t^a$ the combination $-\TT{a}{b}t^b$ is a future directed causal vector.

\noindent
A general spherically symmetric $4$-dimensional spacetime's metric components in the coordinates $(\tau,\ro,\teta,\fii)$ can be written in the form
\begin{equation}
(\dd s_g)^2=-\alpha\beta^2(\dd\tau)^2+\alpha(\dd\ro)^2+r^2\left[(\dd\teta)^2+\sin^2\teta(\dd\fii)^2\right],
\end{equation}
where $\alpha$, $\beta$ and $r$ are functions of $\tau$ and $\ro$. We choose the time orientation as
\begin{equation}
t^a=\frac{1}{\sqrt{\alpha}\beta}(\dd\tau)^a
\end{equation}
is future pointing timelike vector field.
\subsection{$3+1$ and $2+2$ splitting}
Let $\maniii$ be a $3$-dimensional smooth hypersurface with $t^a$ as unit normal. Let $\h{}{ab}$ be the induced Riemannian metric on $\maniii$ defined trough the projection operator
\begin{equation}
\h{a}{b}=\g{a}{b}+t^at_b,
\end{equation}
as
\begin{equation}
\h{}{ab}=\h{e}{a}\h{f}{b}\g{}{ef}.
\end{equation}
The corresponding volume form is denoted by $\vol{h}$.\\
The $SO(3)$ invariant $\manii$ surface is embedded in $\maniii$. Its spacelike unit normal tangential to $\maniii$ is
\begin{equation}
r^a=\frac{1}{\sqrt{\alpha}}(\dd\ro)^a.
\end{equation}
Instead $t^a$ and $r^a$ one may choose lightlike basis vector fields orthogonal to $\manii$. Denote these vector fields with $n_a$ and $\ell_a$, where $n_a$ is past directed outward pointing and $\ell_a$ is future directed outward pointing satisfying $n^a\ell_a=1$.\\
One may be interested in the $\expl$ and $\expn$ null expansions of $\manii$ defined by \cite{unti}\cite{hawki}
\begin{equation}
\pounds_\ell\vol{q}=\expl\vol{q},\qquad\pounds_n\vol{q}=\expn\vol{q},
\end{equation}
Where $\pounds_\ell$ denotes Lie derivative with respect to vector field $\ell^a$ and $\vol{q}$ denotes the volume form induced on $\manii$.\\
The $2$-dimensional spacelike surface $\manii$ with $\expl\expn=0$ is called marginal surface. The surface with $\expl\expn<0$ is called trapped, and $\expl\expn>0$ is called untrapped. A marginal surface is called future if $\expl=0$ and $\expn>0$ hold. In this case, if $\pounds_n\expl<0$, we call the future marginal surface is outer. The future marginal surface with $\pounds_n\expl>0$ is called inner \cite{cao}. For sake of simplicity we only investigate future marginal surfaces and simply refer to them as outer or inner marginal surfaces.
\subsection{Misner-Sharp mass}
In spherically symmetric spacetimes there is a vector field, called Kodama vector, which is divergence free $\nabla_aK^a=0$ and also the construction $\GG{ab}{}\nabla_aK_b$ vanishes \cite{mmass,gombi}. Kodama vector is defined as
\begin{equation}
K^a=\varepsilon^{ab}\nabla_b r,
\end{equation}
where $\varepsilon^{ab}$ is the volume form of the submanifold orthogonal to $\manii$. Using this vector field one can define a locally conserved energy-current, called Kodama current:
\begin{equation}
J^a=\frac{1}{8\pi}K^b\GG{a}{b}.
\end{equation}
Misner-Sharp mass is defined as \cite{gombi}
\begin{equation}
\label{eq:msm1}
M=\frac{r}{2}\left(1-\g{ab}{}\nabla_{a}r\nabla_{b}r\right).
\end{equation}
Equation (\ref{eq:msm1}) may be rephrased as \cite{gombi}:
\begin{eqnarray}
\label{eq:msm2}
M=\frac{r}{2}\left(1-\frac{r^2}{2}\expn\expl\right),\\
\label{eq:msm3}
M=m_0+\int_{\ball}J^at_a\vol{h}.
\end{eqnarray}
Since the integrand in (\ref{eq:msm3}) is non-negative if DEC holds and $K^a$ causal, Misner-Sharp mass is guaranteed to be positive if there are no trapped surfaces inside $\manii$. Misner-Sharp mass is also positive for trapped surfaces since $\expn\expl<0$. This means $M$ is non-negative if DEC holds.\\
The equation in the center of our investigation is the combination of (\ref{eq:msm2}) and (\ref{eq:msm3}) reads as:
\begin{equation}
\frac{r^2}{2}\expn\expl=1-\frac{2}{r}\left(m_0+\int_{\ball}J^at_a\vol{h}\right).
\end{equation}

\section{Electromagnetic fields}
\label{sec:em}
Electromagnetic fields are represented by the Faraday tensor $\FF{}{ab}$. This tensor can be expressed in terms of $E^a$ electric and $H^a$ magnetic fields as \cite{maxi,max2}
\begin{equation}
\FF{}{ab}=2t_{[a}E_{b]}+\varepsilon_{abcd}t^cH^d
\end{equation}
where
\begin{equation}
E^a=\h{a}{b}E^b\quad\mathrm{and}\quad H^a=\h{a}{b}H^b
\end{equation}
hold. On $\maniii$ these fields have to satisfy the Maxwell constraint equations:
\begin{equation}
\label{eq:maxconst}
\DD_aE^a+2\omega_aH^a=q\quad\mathrm{and}\quad\DD_aH^a-2\omega_aE^a=0,
\end{equation}
where $\omega_a$ is the twist vector, $\DD_a$ is the covariant derivative operator associated with $\h{}{ab}$ and $q$ is electric charge density \cite{max2}. Electric charge contained in $\ball$ is defined by the integral
\begin{equation}
Q=\int_\ball q\vol{h}.
\end{equation}
For spherically symmetric spacetimes $\omega_a=0$ and the only non-vanishing component of $E^a$ and $H^a$ is the normal one to $\manii$. Using this fact (\ref{eq:maxconst}) gives the following solutions:
\begin{equation}
\label{eq:maxsol}
E=\frac{Q}{r^2},\qquad H=0,
\end{equation}
where $E=\sqrt{E_aE^a}$ and $H=\sqrt{H_aH^a}$ similarly.\\
The stress-energy tensor for electromagnetic fields is defined by
\begin{equation}
\TT{}{ab}=\frac{1}{4\pi}\left[-\FF{}{ac}\FF{c}{b}-\frac{\g{}{ab}}{4}\FF{ef}{}\FF{}{ef}\right]
\end{equation}
which can be expressed using $E^a$ and $H^a$ as \cite{max2}
\begin{equation}
\label{eq:emes2}
\TT{}{ab}=\frac{1}{4\pi}\left[\frac{1}{2}(E^2+H^2)t_at_b+\frac{1}{6}(E^2+H^2)\h{}{ab}+2S_{(a}t_{b)}+P_{ab}\right].
\end{equation}
In (\ref{eq:emes2}) $S_a=\epsilon_{abcd}E^bH^ct^d$ is the Poynting vector which vanishes in case of spherically symmetric spacetimes and
\begin{equation}
P_{ab}=\frac{1}{3}(E^2+H^2)\h{}{ab}-E_aE_b-H_aH_b.
\end{equation}
Then using (\ref{eq:maxsol}) and (\ref{eq:emes2}) we get
\begin{equation}
J^at_a=\frac{1}{8\pi\sqrt{\alpha}}\frac{Q^2}{r^4}\diff_\ro r.
\end{equation}

\section{Applications}
\label{sec:app}
We use two physical settings to derive inequalities. The first one is a charged body surrounded by electrovacuum. There are two possibilities: if marginal surface occur outside the body or not. In the former case we study this marginal surface, in the latter case we study the body itself. Studying the marginal surface we get inequalities for inner marginal surface. One might ask if these inequalities are applicable to wide classes of inner marginal surfaces as for black holes. To test this idea we shortly mention the case of flat FLRW spacetime.
\subsection{A body and its mass}
\label{sec:mass}
\begin{figure}[ht]
 \includegraphics[width=\textwidth]{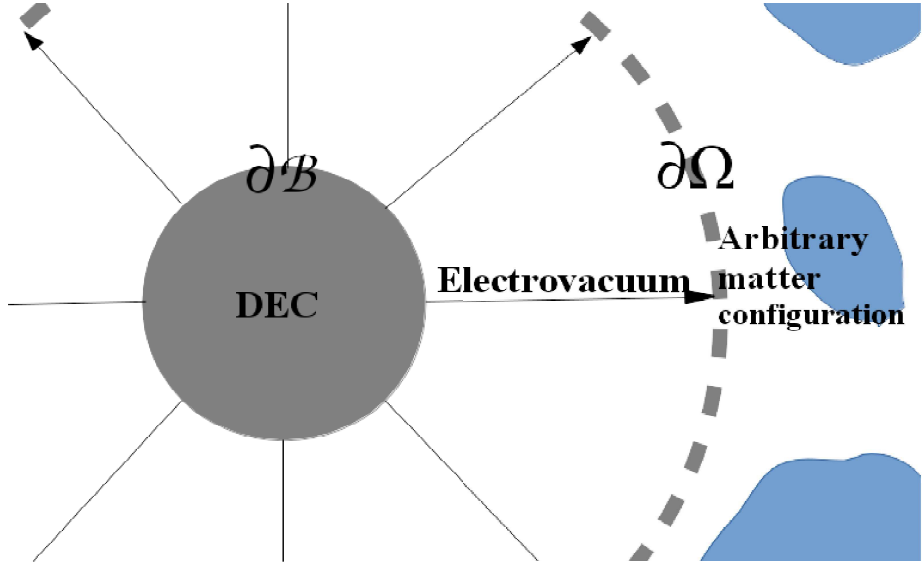}
 \caption{The geometrical setup for a charged body. The gray sphere symbolize the body. In this gray sphere the dominant energy condition holds. Between the sphere and the dashed lines only gravity and electromagnetism acts to distinguish the body from other parts of spacetime. Outside the dashed lines matter is distributed arbitrarily.}
 \label{fig:geo}
\end{figure}

The spherically symmetric charged body is modeled by an origin centered $\ball$ ball with areal radius $r_0$ surrounded by electrovacuum as the only long range interaction. Let $\world$ be the ball in which measurements was taken such that $\ball\subset\world$. Let take $\manii$ surface as $\manii=\partial\world$ with areal radius $r$! It is assumed that matter in $\ball$ satisfy DEC. There are no constraints regarding the matter fields outside $\world$. This geometric setup is indicated in figure \ref{fig:geo}.\\
Then we have for the body's mass
\begin{equation}
m^*=m_0+\int_\ball J^at_a\vol{h}.
\end{equation}
The mass contained in the ball bounded by $\manii$ is
\begin{equation}
\label{eq:m1}
M=m^*+\int_{\world\backslash\ball}\frac{1}{8\pi\sqrt{\alpha}}\frac{Q^2}{r^4}\diff_\ro r\vol{h}=m^*+\frac{Q^2}{2r_0}-\frac{Q^2}{2r}
\end{equation}
since $Q$ is constant in electrovacuum regions. We introduce the notation
\begin{equation}
\label{eq:m2}
m=m^*+\frac{Q^2}{2r_0}.
\end{equation}
Note that $m$ may be a better mass notion than $m^*$ in the following sense. Let see the case when the entire spacetime is electrovacuum outside $\ball$ and the metric (in particular $\alpha(\tau,\ro)$) is asymptotically flat. In this case the spacetime is asymptotically flat so we can compare $m$ and $m^*$ to the ADM mass. It is known that Misner-Sharp mass is equal to ADM mass in the limit $r\to\infty$ \cite{quasjar}. Taking the limit of (\ref{eq:m1}) we get that the one which recover the ADM mass is $m$.\\
The Misner-Sharp mass and the product of expansions in the case introduced above read as
\begin{eqnarray}
M=m-\frac{Q^2}{2r},\\
\label{eq:chex}
\frac{r^2}{2}\expl\expn=1-\frac{2m}{r}+\frac{Q^2}{r^2}.
\end{eqnarray}
The following two applications use these equations to study marginal surfaces and charged bodies in electrovacuum.
\subsection{Marginal surfaces in electrovacuum}
\label{sec:msem}
As geometric inequalities are important tools to understand the evolution of black holes the first subject of application is a charged black hole. However the vanishing of $\expn\expl$ do not distinguish between marginally inner or outer trapped surfaces. This way one may get relations similar as in \cite{Hennig2009} for the developing Cauchy horizon.\\
It follows from (\ref{eq:chex}) that marginal surfaces may occur at
\begin{equation}
r_\pm=m\pm\sqrt{m^2-Q^2}
\end{equation}
if $|Q|<m$ and either $r_-$ or $r_+$ is in $\world\backslash\ball$. We note that the matter distribution outside $\world$ may prevent the existence of marginal surfaces if neither $r_-$ nor $r_+$ is in $\world$. In case there is a marginal surface the expansion $\expl$ changes as
\begin{equation}
\pounds_n\expl=\frac{\RR{}{q}}{2}+\GG{}{ab}n^al^b=\frac{1}{r^2}-\frac{Q^2}{r^4}.
\end{equation}
For $r_+$ $\pounds_n\expl>0$ thus it is outer marginal surface while for $r_-$ $\pounds_n\expl<0$ thus it is inner marginal surface.\\
In the first case one get the well known \cite{Dain:2011kb,Khuri:2015xpa,Anglada:2015tan} inequality between size and charge. Using 
\begin{eqnarray}
\label{eq:+1}
r=r_+\\
\label{eq:+2}
|Q|\leq m
\end{eqnarray}
yield
\begin{equation}
\label{eq:+AQ}
\frac{A_+}{4\pi}\geq Q^2,
\end{equation}
%
%
%
with equality in the extreme case. This may be considered as the spherically symmetric version of (\ref{eq:AJQ}). (\ref{eq:+1}) and (\ref{eq:+2}) also implying that
\begin{equation}
m^2\in\left[\frac{A_+}{16\pi};\frac{A_+}{4\pi}\right].
\label{eq:ma+}
\end{equation}
$A_+=4\pi m^2$ is reached in extremal case ($Q^2=m^2$), $A_+=16\pi m^2$ is reached in the vacuum case ($Q^2=0$). The inequality of the global version of
\begin{equation}
\frac{A_+}{16\pi}\leq m^2
\end{equation}
 is known as Penrose inequality \cite{geosumm}. We recovered these inequalities without assuming asymptotic flatness, provided that the relation (\ref{eq:ma+}) holds.\\
In case of inner marginal surfaces using 
\begin{eqnarray}
r=r_-\\
|Q|\leq m
\end{eqnarray}
yields
\begin{eqnarray}
\label{eq:-AQ}
\frac{A_-}{4\pi}\leq Q^2,\\
\label{eq:-Am}
\frac{A_-}{4\pi}\leq m^2,
\end{eqnarray}
both saturated in the extremal case. The inequality (\ref{eq:-AQ}) is known \cite{Hennig2009} in the form
\begin{equation}
A_+A_-=(8\pi Q^2)^2.
\end{equation}
%
%
\subsection{Spherically symmetric charged body}
\label{sec:chb}
As it was argued in \cite{Khuri:2015xpa} geometric inequalities for bodies may implicate criteria for black hole forming. Based on the previous results of section \ref{sec:msem} in case of a charged body if it is surrounded by untrapped surfaces either of the following properties hold:
\begin{itemize}
\renewcommand{\labelitemi}{$\circ$}
\item $|Q|<m$ and $r_0<r_-$ or
\item $|Q|<m$ and $r_0>r_+$ or
\item $|Q|>m$.
\end{itemize}
The first and second case is similar to the marginal surface's case using $A_0<A_-$ or $A_0>A_+$. The importance of the second is clear but the first may hold only with some unstable matter distribution outside $\world$ to avoid forming trapped surface.  The result is
\begin{equation}
\frac{A}{4\pi}>Q^2\quad\mathrm{and}\quad\frac{A}{4\pi}>m^2\quad\mathrm{for}\quad r_0>r_+
\end{equation}
and
\begin{equation}
\frac{A}{4\pi}<Q^2\quad\mathrm{and}\quad\frac{A}{4\pi}<m^2\quad\mathrm{for}\quad r_0<r_-.
\end{equation}
In the third case one can use the inequality
\begin{equation}
m-\frac{Q^2}{2r}>0
\end{equation}
which is equivalent to (\ref{eq:reiris}).
%
%
%
%
Except in case $|Q|<m$ and $r_0<r_-$ the inequality between $A$ and $Q^2$ is different only in a positive constant from (\ref{eq:+AQ}) as proved in \cite{Khuri:2015xpa,Anglada:2015tan}. However in case $|Q|<m$ and $r_0<r_-$ the difference is a negative constant.

\subsection{Homogeneous isotropic cosmology}
\label{sec:cosm}
We have seen in \ref{sec:chb} that inner horizons may be investigated by the proposed method. The inner horizon of charged black hole may have less physical relevance in the sense it is unlikely that such an object is formed in dynamical processes. A more relevant context where inner horizons play a role is the cosmological horizon in cosmological models. In spite of the fact that the presence of electric charge is not considered to the relevant cosmological investigations here we include it for sake of completeness. 
An FLRW spacetime is spherically symmetric around any of its events thereby everything we stated in section \ref{sec:prelim} is also applicable to FLRW spacetimes. 
In this case using the following substitutions: $\tau \to t;\, \ro \to \tilde{r};\, \alpha\to a(t)^2;\, \beta\to 1/a(t);\, r\to a(t)\tilde{r}$
one gets the line element of a flat FLRW spacetime:
\begin{equation}
(\dd s)^2=-(\dd t)^2+a(t)^2\left[(\dd\tilde{r})^2+\tilde{r}^2((\dd\teta)^2+\sin^2\teta(\dd\fii)^2)\right].
\end{equation}
Mass and charge can be expressed via integrals of homogeneous isotropic densities $\rho$ and $q$ as:
\begin{equation}
m=\rho\frac{4\pi r^3}{3}\qquad\mathrm{and}\qquad Q=q\frac{4\pi r^3}{3}.
\end{equation}
Using these densities and taking account of cosmological constant the product of expansions reads as
\begin{equation}
\frac{r^2}{2}\expl\expn=1-(8\pi\rho+\Lambda)\frac{r^2}{3}+\frac{16\pi^2}{9}q^2r^4.
\end{equation}
If $8\pi\rho+\Lambda\geq 8\pi q$ there are marginal surfaces at the location
\begin{equation}
r_\pm^2=\frac{3}{32\pi^2q^2}\left[(8\pi\rho+\Lambda)\pm\sqrt{(8\pi\rho+\Lambda)^2-64\pi^2q^2}\right],
\end{equation}
and the following inequalities can be seen to hold:
\begin{eqnarray}
\label{eq:+Aq}
A_+\geq \frac{3}{q}\qquad\mathrm{or}\qquad Q^2\geq \frac{A_+}{4\pi};\\
\label{eq:-Aq}
A_-\leq \frac{3}{q}\qquad\mathrm{or}\qquad Q^2\leq \frac{A_-}{4\pi}.
\end{eqnarray}
Clearly (\ref{eq:+Aq}) may not have a deep physical relevance but it may be interesting to compare with the inequalities obtained in section \ref{sec:msem}. Inequality (\ref{eq:-Aq}) corresponds to (\ref{eq:+AQ}) with the difference that it applies to an inner marginal surface in the present case.

\section{Summary}
\label{sec:summ}

The quasi-local concept of Misner-Sharp mass was used to investigate geometric inequalities in spherically symmetric spacetimes. Misner-Sharp mass is related to the null expansions of surfaces so it is a useful concept to distinguish marginal, trapped or untrapped surfaces. Using this feature we derived the known inequalities for black holes and normal bodies without any assumption on the asymptotic behavior in spherical symmetric spacetimes. It is important to be emphasized that the results suggest that two bodies surrounded by untrapped surfaces may be very different. Some of our foundings may be important in a better modeling of normal bodies.\\
Applying our method we derived a quasi-local version of Penrose inequality and quasi-local relations between mass, surface area and charge, which can be used to derive quasi-local versions of black hole dynamic laws.
Similarly bounds on mass of normal bodies was established.\\
In two particular cases the presence of inner marginal surfaces were also studied. We found that geometric inequalities between its area and charge is not characteristic to these kind of surfaces. In the first one we got the relation $A\leq 4\pi Q^2$, however, in homogeneous isotropic spacetime we got the opposite $A\geq 4\pi Q^2$. It is not clear what is the reason beyond these differences and if any kind of definite inequalities may hold for certain class of inner marginal surfaces.\\
We found that using quasi-local mass concept may be fruitful in these kind of investigations. We wish to apply our method to more generic spacetimes with fewer symmetries.

\section{Acknowledgements}
I would like to thank I. R\'{a}cz for suggesting the problem and for illuminating discussions.

\bibliographystyle{plain}

\bibliography{bibi.bib}   

\begin{thebibliography}{10}

\bibitem{Anglada:2015tan}
Pablo Anglada, Sergio Dain, and Omar~E. Ortiz.
\newblock {Inequality between size and charge in spherical symmetry}.
\newblock {\em Phys. Rev.}, D93(4):044055, 2016.

\bibitem{Ansorg:2005bq}
Marcus Ansorg and David Petroff.
\newblock {Black holes surrounded by uniformly rotating rings}.
\newblock {\em Phys. Rev.}, D72:024019, 2005.

\bibitem{Ansorg:2007fh}
Marcus Ansorg and Herbert Pfister.
\newblock {A Universal constraint between charge and rotation rate for
  degenerate black holes surrounded by matter}.
\newblock {\em Class. Quant. Grav.}, 25:035009, 2008.

\bibitem{Bertotti:1959pf}
B.~Bertotti.
\newblock {Uniform electromagnetic field in the theory of general relativity}.
\newblock {\em Phys. Rev.}, 116:1331, 1959.

\bibitem{cao}
Li-Ming Cao.
\newblock {Deformation of Codimension-2 Surface and Horizon Thermodynamics}.
\newblock {\em JHEP}, 03:112, 2011.

\bibitem{Chrusciel:2009ki}
Piotr~T. Chrusciel and Joao Lopes~Costa.
\newblock {Mass, angular-momentum, and charge inequalities for axisymmetric
  initial data}.
\newblock {\em Class. Quant. Grav.}, 26:235013, 2009.

\bibitem{Clement:2012vb}
Maria E.~Gabach Clement, Jose~Luis Jaramillo, and Martin Reiris.
\newblock {Proof of the area-angular momentum-charge inequality for
  axisymmetric black holes}.
\newblock {\em Class. Quant. Grav.}, 30:065017, 2013.

\bibitem{gombi}
P.~Csizmadia and I.~R\'{a}cz.
\newblock {Gravitational collapse and topology change in spherically symmetric
  dynamical systems}.
\newblock {\em Classical and Quantum Gravity}, 27:015001, 2010.

\bibitem{Dain:2005qj}
Sergio Dain.
\newblock {Proof of the (local) angular momemtum-mass inequality for
  axisymmetric black holes}.
\newblock {\em Class. Quant. Grav.}, 23:6845--6856, 2006.

\bibitem{geosumm}
Sergio Dain.
\newblock {\em {Geometric Inequalities for Black Holes}}, volume 157, pages
  51--52.
\newblock 2014.

\bibitem{Dain:2011kb}
Sergio Dain, Jose~Luis Jaramillo, and Martin Reiris.
\newblock {Area-charge inequality for black holes}.
\newblock {\em Class. Quant. Grav.}, 29:035013, 2012.

\bibitem{Dain:2011pi}
Sergio Dain and Martin Reiris.
\newblock {Area - Angular momentum inequality for axisymmetric black holes}.
\newblock {\em Phys. Rev. Lett.}, 107:051101, 2011.

\bibitem{mmass}
S.~A. Hayward.
\newblock {Gravitational energy in spherical symmetry}.
\newblock {\em Physical Review}, D53:1938--1949, 1996.

\bibitem{Hennig2009}
Jorg Hennig and Marcus Ansorg.
\newblock {The Inner Cauchy horizon of axisymmetric and stationary black holes
  with surrounding matter in Einstein-Maxwell theory: Study in terms of soliton
  methods}.
\newblock {\em Annales Henri Poincare}, 10:1075--1095, 2009.

\bibitem{Hennig:2008yw}
Jorg Hennig, Marcus Ansorg, and Carla Cederbaum.
\newblock {A Universal inequality between angular momentum and horizon area for
  axisymmetric and stationary black holes with surrounding matter}.
\newblock {\em Class. Quant. Grav.}, 25:162002, 2008.

\bibitem{Hennig:2008zy}
Jorg Hennig, Carla Cederbaum, and Marcus Ansorg.
\newblock {A Universal inequality for axisymmetric and stationary black holes
  with surrounding matter in the Einstein-Maxwell theory}.
\newblock {\em Commun. Math. Phys.}, 293:449--467, 2010.

\bibitem{quasjar}
J.~L. Jaramillo and E.~Gourgoulhon.
\newblock {Mass and Angular Momentum in General Relativity}.
\newblock {\em Fundamental Theories of Physics}, 162:87--124, 2011.
\newblock [,87(2010)].

\bibitem{Jaramillo:2011pg}
Jose~Luis Jaramillo, Martin Reiris, and Sergio Dain.
\newblock {Black hole Area-Angular momentum inequality in non-vacuum
  spacetimes}.
\newblock {\em Phys. Rev.}, D84:121503, 2011.

\bibitem{Khuri:2015xpa}
M.~A. Khuri.
\newblock {Inequalities Between Size and Charge for Bodies and the Existence of
  Black Holes Due to Concentration of Charge}.
\newblock {\em Journal of Mathematical Physics}, 56(11):112503, 2015.

\bibitem{Khuri:2015zla}
Marcus~A. Khuri.
\newblock {Existence of Black Holes Due to Concentration of Angular Momentum}.
\newblock {\em JHEP}, 06:188, 2015.

\bibitem{unti}
I.~R\'{a}cz.
\newblock {On the topology of untrapped surfaces}.
\newblock {\em Classical and Quantum Gravity}, 26:055017, 2009.

\bibitem{maxi}
Istvan Racz.
\newblock {Maxwell fields in space-times admitting nonnull killing vectors}.
\newblock {\em Class. Quant. Grav.}, 10:L167--L172, 1993.

\bibitem{hawki}
Istvan Racz.
\newblock {A Simple proof of the recent generalisations of Hawking's black hole
  topology theorem}.
\newblock {\em Class. Quant. Grav.}, 25:162001, 2008.

\bibitem{Reiris:2014tva}
Martin Reiris.
\newblock {On the shape of bodies in General Relativistic regimes}.
\newblock {\em Gen. Rel. Grav.}, 46:1777, 2014.

\bibitem{Robinson:1959ev}
I.~Robinson.
\newblock {A Solution of the Maxwell-Einstein Equations}.
\newblock {\em Bull. Acad. Pol. Sci. Ser. Sci. Math. Astron. Phys.},
  7:351--352, 1959.

\bibitem{max2}
C.~G. Tsagas.
\newblock {Electromagnetic fields in curved spacetimes}.
\newblock {\em Classical and Quantum Gravity}, 22:393--408, 2005.

\end{thebibliography}

\end{document}